\documentclass{article}
\usepackage{cite}
\usepackage{spconf,algorithm,algorithmic,amsmath,amssymb,amsthm,bbm,color,graphicx,microtype,url}
\usepackage{comment}
\usepackage{subcaption}
\usepackage{xcolor, soul}
\sethlcolor{yellow}

\renewcommand{\a}{\mathbf{a}}

\newcommand{\n}{\mathbf{n}}

\newcommand{\s}{\mathbf{s}}

\newcommand{\x}{\mathbf{x}}
\newcommand{\y}{\mathbf{y}}

\newcommand{\D}{\mathbf{D}}

\renewcommand{\H}{\mathbf{H}}
\newcommand{\I}{\mathbf{I}}

\renewcommand{\P}{\mathbf{P}}

\newcommand{\U}{\mathbf{U}}
\newcommand{\V}{\mathbf{V}}
\newcommand{\W}{\mathbf{W}}

\newcommand{\Compl}{\mbox{$\mathbb{C}$}}

\title{Joint Near-Field Target Tracking and Communications\\ with Full Duplex Holographic MIMO}
\name{Ioannis Gavras and George~C.~Alexandropoulos
\thanks{This work was supported by the Smart Networks and Services Joint Undertaking (SNS JU) project TERRAMETA under the European Union’s Horizon Europe research and innovation programme under Grant Agreement No 101097101, including top-up funding by UK Research and Innovation
(UKRI) under the UK government’s Horizon Europe funding guarantee.}}
\address{Department of Informatics and Telecommunications, National and Kapodistrian University of Athens,\\
Panepistimiopolis Ilissia, 15784 Athens, Greece}

\begin{document}
\maketitle

\begin{abstract}
In this paper, we present a simultaneous target tracking and multi-user communications system realized by a full duplex holographic Multiple-Input Multiple-Output (MIMO) node equipped with Dynamic Metasurface Antennas (DMAs) at both its communication ends. Focusing on the near-field regime, we extend Fresnel's approximation to metasurfaces and devise a subspace tracking scheme with DMA-based hybrid Analog and Digital (A/D) reception as well as hybrid A/D transmission with a DMA for sum-rate maximization. The presented simulation results corroborate the efficiency of the proposed framework for various system parameters.
\end{abstract}


\section{Introduction}
In-Band Full Duplex (FD) Multiple-Input Multiple-Output (MIMO) systems~\cite{sabharwal2014band,alexandropoulos2017joint,Iimori2019,FD_AUXTX,FD_MIMO_Arch,FD_survey,FD_ISAC_Asilomar} are lately being investigated as an efficient technology for Joint Communications And Sensing (JCAS)~\cite{liu2022integrated,mishra2019toward}, which constitutes a key connectivity paradigm for sixth Generation (6G) wireless networks. The main challenge for such simultaneous Transmit (TX) and Receive (RX) operations is Self Interference (SI) appearing at the FD node, which can become more severe as the number of TX and RX antenna increases. To deal with this, solutions based on TX/RX isolation, Analog and Digital (A/D) SI cancellation, and hybrid A/D BeamForming (BF) have been presented \cite{Vishwanath_2020, alexandropoulos2020full,islam2021direction,FD_MIMO_VTM2022,WB_FD_MIMO_TWC2022}. 

Prior research on FD-enabled JCAS systems ranges from single-antenna schemes~\cite{barneto2019full,liyanaarachchi2021optimized} to massive MIMO under millimeter-wave frequencies~\cite{barneto2020beamforming,Islam_2022_ISAC,Atiq_ISAC_2022}. Very recently, \cite{gavras2023full} capitalized on the holographic MIMO concept realized by TX/RX Dynamic Metasurface Antenna (DMA) arrays~\cite{HMIMO_survey} and studied simultaneous data communications and target localization in the challenging THz frequency band and the near-field regime. The super focusing capability of massive DMAs was leveraged to efficient treat SI, while enabling communications in the DownLink (DL) and a variant of MUltiple SIgnal Classification (MUSIC) for locazation in the uplink. However, efficient target tracking with such extremely massive FD MIMO systems has not been yet investigated.

In this paper, we capitalize on \cite{gavras2023full}'s FD-enabled JCAS framework and present a simultaneous target tracking and multi-user communications system. By extending Fresnel's approximation~\cite{pan2022ris} to Uniform Planar Arrays (UPAs), and consequently DMAs, we leverage the reflected DL signals from mobile targets to achieve precise tracking of their parameters, while maximizing DL communications. For this goal, an optimization framework for the joint design of TX/RX A/D BF and the digital SI cancellation is presented. Our simulation results demonstrate the effectiveness of our approach, revealing intricate interactions among various system parameters.

\textit{Notations:} Boldface lowercase and boldface capital letters represent vectors and matrices, respectively. $\mathbf{A}^{\rm T}$, $\mathbf{A}^{\rm H}$, $[\mathbf{A}]_{i,j}$, and $\|\mathbf{A}\|$ denote $\mathbf{A}$'s transpose, Hermitian transpose, $(i,j)$th element, and Euclidean norm, respectively. $\mathbb{C}$ is the complex number set and $\jmath$ is the imaginary unit. $\mathbb{E}\{\cdot\}$ is the expectation operator and $\mathbf{x}\sim\mathcal{CN}(\mathbf{a},\mathbf{A})$ indicates a complex Gaussian random vector with mean $\mathbf{a}$ and covariance matrix $\mathbf{A}$.

\section{System and Signal Models}\label{sec: system_signal}
\subsection{FD Holographic MIMO JCAS System Model}
A DMA-based FD holographic MIMO transceiver, similar to \cite[Fig. 1]{gavras2023full}, wishing to communicate in the DL direction with $U$ $L$-antenna Users' Equipment (UEs), while simultaneously receiving in the uplink the reflections of its transmitted signals from $K\geq U$ targets in its vicinity is considered. The $U$ served UEs belong to this set of $K$ sensed targets, while the remaining $K-U$ targets are either passive objects or non-cooperating UEs. An approach similar to~\cite{RISslam} to separate the reflections from the $U$ served UEs and $K-U$ targets is assumed. The TX/RX DMAs are located in the $xz$-plane, with each comprising $N_{\rm RF}$ microstrips of $N_{\rm E}$ metamaterials each, where each microstrip is connected to a TX/RX Radio Frequency (RF) chain ($N_{\rm RF}>L$ in total). The inter-element distance within each microstrip is $d_{\rm E}$ and the horizontal separation between TX and RX DMAs is $d_{\rm pl} = 2d_{\rm P}$. Consequently, both arrays have a total of $N \triangleq N_{\rm RF}N_{\rm E}$ metamaterials. All $U$ UEs requesting DL communications are equipped with an $L$-element fully digital Uniform Linear Array (ULA) situated along the $z$-axis.

Let the $N\times N$ diagonal matrices $\P_{\rm TX}$ and $\P_{\rm RX}$, whose elements model the signal propagation inside the microstrips at the TX and RX DMAs, respectively. The former is defined $\forall$$i=1,2,\ldots,N_{\rm RF}$ and $\forall$$n = 1,2,\ldots,N_{\rm E}$ as \cite{Xu_DMA_2022}:
\begin{align}\label{eq: TX_Sig_Prop}
    [\P_{\rm TX}]_{(i-1)N_{\rm E}+n,(i-1)N_{\rm E}+n} \!\triangleq\! \exp{(-\rho_{i,n}(\alpha_i + \jmath\beta_i))},
\end{align}
where $\alpha_i$ is the waveguide attenuation coefficient, $\beta_i$ is the wavenumber, and $\rho_{i,n}$ denotes the location of the $n$th element in the $i$th microstrip. Similar is the definition for $\P_{\rm RX}$. Let $w^{\rm TX}_{i,n}$ and $w^{\rm RX}_{i,n} $ denote the tunable responses of the TX/RX DMAs, respectively, for each $n$th metamaterial and each $i$th microstrip, which are assumed to follow a Lorentzian-constrained phase model and belong to the phase profile codebook:
\begin{align}
    \mathcal{W}\triangleq \left\{0.5\left(\jmath+e^{\jmath\phi}\right)\Big|\phi\in\left[-\frac{\pi}{2},\frac{\pi}{2}\right]\right\}.
\end{align}
The analog TX BF matrix  $\W_{\rm TX}\in\mathbb{C}^{N\times N_{\rm RF}}$ is given by:
\begin{align}
    [\W_{\rm TX}]_{(i-1)N_{\rm E}+n,j} = \begin{cases}
    w^{\rm TX}_{i,n},&  i=j,\\
    0,              & i\neq j.
\end{cases}
\end{align}
Similarly, we define the weights $w^{\rm RX}_{i,n} \in \mathcal{W}$ $\forall$$i,n$ from which the analog combining matrix $\W_{\rm RX}\in\mathbb{C}^{N\times N_{\rm RF}}$ is formulated.

The DMA-based TX of the considered FD transceiver possesses the symbol vector $\s_{u}\in\Compl^{L\times1}$ for each $u$th UE ($u=1,2,\ldots,U$), which is precoded via the digital matrix $\V_u\in\Compl^{N_{\rm RF}\times L}$. Before DL transmission, the digitally precoded symbols are analog processed via the weights of the TX DMA, yielding the $N$-element transmitted signal $\x\triangleq\P_{\rm TX}\W_{\rm TX}\V\s$ where $\V \triangleq [\V_1,\V_2,\ldots,\V_U]\in\Compl^{N_{\rm RF}\times UL}$ and $\s \triangleq [\s_1^{\rm T},\s_2^{\rm T},\ldots,\s_U^{\rm T}]^{\rm T}\in\Compl^{UL\times1}$. We finally assume this signal is power limited such that $\mathbb{E}\{\|\P_{\rm TX}\W_{\rm TX}\V\s\|^2\}\leq P_{\rm max}$ with $P_{\rm max}$ indicating the maximum transmission power.

\subsection{Near-Field Channel Model}
The $L\times N$ complex-valued DL channel between each $u$th UE and the DMA-based TX of the considered FD holographic MIMO transceiver is modeled as follows:
\begin{align}
    \label{eqn:DL_chan}
    [\H_{{\rm DL},u}]_{\ell,(i-1)N_{\rm E}+n} \triangleq \alpha_{u,\ell,i,n} \exp\left(\frac{\jmath2\pi}{\lambda} r_{u,\ell,i,n}\right),
\end{align}
where $r_{u,\ell,i,n}$ represents the distance between the $\ell$th antenna ($\ell=1,2,\ldots,L$) of each $u$th UE and the $n$th metamaterial of each $i$th TX microstrip. In addition, $\alpha_{u,\ell,i,n}$ denotes the respective attenuation factor with $\kappa_{\rm abs}$  being the molecular absorption coefficient at THz, which is defined as:
\begin{align}\label{eq: atn}
    \alpha_{u,\ell,i,n} \triangleq \sqrt{F(\theta_{u,\ell,i,n})}\!\frac{\lambda}{4\pi r_{u,\ell,i,n}}\!\exp\left(-\frac{\kappa_{\rm abs}r_{u,\ell,i,n}}{2}\right)
\end{align}
with $\lambda$ being the wavelength and $F(\cdot)$ is each metamaterial's radiation profile, modeled for an elevation angle $\theta$ as follows:
\begin{align}
    F (\theta) = \begin{cases}
    2(b+1)\cos^{b}(\theta),& {\rm if}\, \theta\in[-\frac{\pi}{2},\frac{\pi}{2}]\\
    0,              & {\rm otherwise}
\end{cases}.
\end{align}
In the latter expression, $b$ determines the boresight antenna gain which depends on the specific DMA technology. 

\begin{figure}[!t]
	\begin{center}
	\includegraphics[width=0.9\columnwidth]{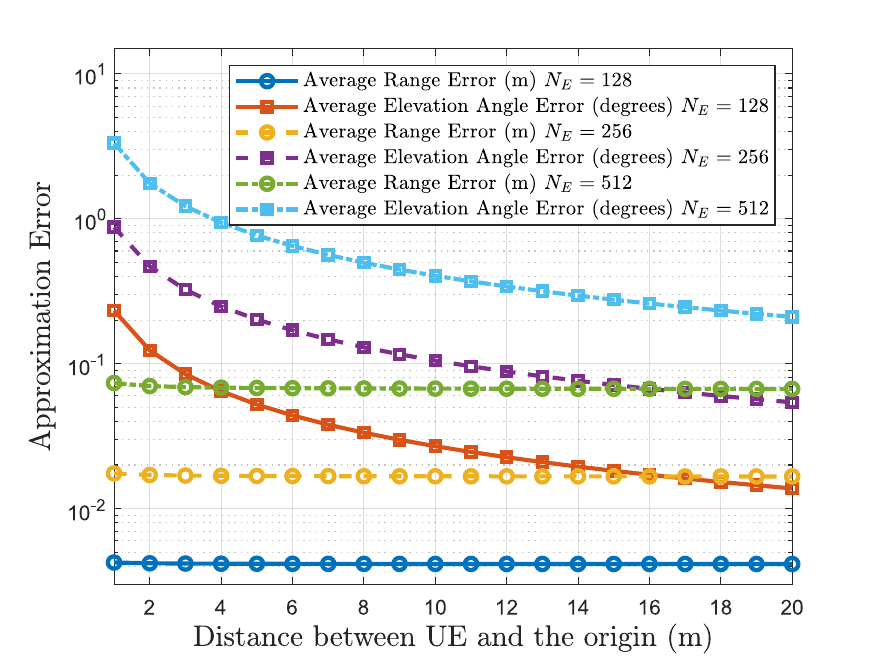}\vspace{-0.2cm}
	\caption{\small{Performance of the $r_{u,\ell,i,n}$ and $\theta_{u,\ell,i,n}$ approximations considering a UE with $L=2$ antennas and different values for $N_{E}$.}}\vspace{-0.4cm}  
	\label{fig: Error}
	\end{center}
\end{figure}

The spherical coordinates of all $K$ targets in the system are the $(r_k,\theta_k,\varphi_k)$ $\forall k=1,2,\ldots,K$, including the distances from the origin, and the elevation and azimuth angles, respectively.  Recall that $U$ out of the $K$ targets with coordinates $(r_u,\theta_u,\varphi_u)$ $\forall u=1,\ldots,U$ are the DL UEs. Each distance $r_{u,\ell,i,n}$ in \eqref{eqn:DL_chan} and \eqref{eq: atn} is given by:
\begin{align}\label{eq: dist}
    \nonumber &r_{u,\ell,i,n}\! =\! \!\Big(\!(r_{u,\ell}\sin\theta_{u,\ell}\cos\varphi_{u,\ell} \!+\!\frac{d_{\rm P}}{2}\!+\!(i\!-\!1)d_{\rm RF})^2 +\\ &(r_{u,\ell}\sin\theta_{u,\ell}\sin\varphi_{u,\ell})^2 \!+\!(r_{u,\ell}\cos\theta_{u,\ell}\!-\!(n\!-\!1)d_{\rm E})^2\Big)^{\frac{1}{2}}\!,
\end{align}
where $r_{u,\ell}$, $\theta_{u,\ell}$, and $\varphi_{u,\ell}$ represent the distance, and the elevation and azimuth angles of each $u$th UE's $\ell$th antenna with respect to the origin; these are computed as in \cite[eq. (8)]{gavras2023full}. We next simplify \eqref{eq: dist} by expressing it as $r_{u,\ell,i,n}(x,z) = r_{u,\ell}F(x,z)$, where function $F(x,z)$ is defined as follows:
\begin{align}\label{eq: Dist_F}
 \nonumber F(x,z)\! = \! &\Big(\!\left(\sin\theta_{u,\ell}\cos\varphi_{u,\ell} \!+\!x\right)^2\!+\\ &(\sin\theta_{u,\ell}\sin\varphi_{u,\ell})^2\!+\! \left(\cos\theta_{u,\ell}\!-\!z\right)^2\Big)^{\frac{1}{2}}.
\end{align}
In this expression, $x = \frac{\frac{d_{\rm P}}{2}+(i\!-\!1)d_{\rm RF}}{r_{\rm u,\ell}}$ and $z = \frac{(n\!-\!1)d_{\rm E}}{r_{\rm u,\ell}}$  
 $\forall$$i,n$. By extending the Fresnel approximation from ULAs to UPA setups, since directly applying it to UPAs is infeasible~\cite{pan2022ris}, the following approximation can be derived: 
%\begin{align}\label{eq: Fresnel_approx}
%    \nonumber &F(x,z)\approx F(x,z)|_{\substack{x=0\\z=0}}+\frac{\partial F}{\partial x}|_{x=0}\times x+
%    \frac{\partial F}{\partial z}|_{z=0}\times z\\\nonumber&+\frac{1}{2}\left(\frac{\partial^2 F}{\partial x^2}|_{\substack{x=0\\z=0}}\times x^2+2\frac{\partial^2 F}{\partial z\partial x}|_{\substack{x=0\\z=0}} + \frac{\partial^2 F}{\partial z^2}|_{\substack{x=0\\z=0}}\right)\\   
%    \nonumber &\approx F(0,0) + \frac{\sin\theta\cos\phi}{F(0,z)}x-\frac{\sin\theta}{F(x,0)}z\\\nonumber
%    & +\left(\frac{1}{F(0,0)}-\frac{\sin^2\theta\cos^2\phi}{F^3(0,0)}\right)\frac{x^2}{2} \\&+ \left(\frac{\sin^2\theta\cos\phi}{F^3(0,0)}\right)\frac{xz}{2} +\left(\frac{1}{F(0,0)}-\frac{\sin^2\theta}{F^3(0,0)}\right)\frac{z^2}{2}.
%\end{align}
\begin{align}\label{eq: Fresnel_approx}
    \nonumber &F(x,z)\approx F(0,0) + \frac{\sin\theta\cos\phi}{F(0,z)}x-\frac{\sin\theta}{F(x,0)}z\\\nonumber
    & +\left(\frac{1}{F(0,0)}-\frac{\sin^2\theta\cos^2\phi}{F^3(0,0)}\right)\frac{x^2}{2} \\&+ \left(\frac{\sin^2\theta\cos\phi}{F^3(0,0)}\right)\frac{xz}{2} +\left(\frac{1}{F(0,0)}-\frac{\sin^2\theta}{F^3(0,0)}\right)\frac{z^2}{2}.
\end{align}
Then, the elevation angle of each $u$th UE's $\ell$th antenna with respect to the $n$th element of each $i$th microstrip is given by:
\begin{align}\label{eq:thetas}
    \theta_{u,\ell,i,n}(x,z) \triangleq \sin^{-1}\left({\frac{|z-\cos{\theta_{u,\ell}}|}{F(x,z)}}\right).
\end{align}
We summarize the validity of the latter approximations in Fig.~\ref{fig: Error}. Therein, we have evaluated the average approximation errors across all antennas, RF chains, and metamaterials for the elevation angle and range when there exist a UE equipped with $L=2$ antennas at various near-field distances, considering different numbers of DMA metamaterials. 

The end-to-end channel between the TX and RX metamaterials, including the single-bounch reflections from all $K$ targets when considered as point sources with coordinates $(r_k,\theta_k,\varphi_k)$ $\forall$$k=1,2,\ldots,K$, is expressed as follows: 
\begin{align}\label{eq:H_R}
    \H_{\rm R} \triangleq \sum\limits_{k=1}^{K}\beta_k \a_{\rm RX}(r_k,\theta_k,\varphi_k)\a_{\rm TX}^{\rm H}(r_k,\theta_k,\varphi_k)
\end{align}
with $\beta_k$ representing the complex-valued reflection coefficient for each $k$th radar target, whereas, using \eqref{eq: atn} and the string definition ${\rm str}\triangleq\{{\rm TX},{\rm RX}\}$, $\a_{\rm str}(\cdot)$ is obtained as:
\begin{align}
    \label{eq:response_vec}
    [\a_{\rm str}(r_k,\theta_k,\phi_k)]_{(i-1)N_E+n} \!\triangleq\! a_{k,i,n}\exp\Big({\jmath\frac{2\pi}{\lambda}r_{k,i,n}}\Big).
\end{align}
In this expression, the elevation angle $\theta_{k,i,n}$ and the distance $r_{k,i,n}$ from the origin for each $k$th target are needed to compute $a_{k,i,n}$. The latter is given by the following expression:
\begin{align}
    \nonumber r_{k,i,n}\triangleq\Big(\big(r_k\sin{\theta_k}\cos{\phi_k}\pm\frac{d_{\rm P}}{2}\pm(i-1)d_{\rm RF}\big)^2\\+\big(r_k\sin{\theta_k}\sin{\phi_k}\big)^2
    +\big(r_k\cos{\theta_k}-(n-1)d_{\rm E}\big)^2\Big)^{\frac{1}{2}}.
\end{align}
The positive sign refers to the TX vector, while the negative sign indicates the RX vector. To simplify $r_{k,i,n}$ and $\theta_{k,i,n}$, one can again apply the two-dimensional Fresnel approximations in \eqref{eq: Fresnel_approx} and \eqref{eq:thetas}, respectively, defining $x$ accordingly.

\subsection{Received Signal Models}
The baseband received signal $\y_{u}\in\Compl^{L\times1}$ at each $u$th UE can be mathematically expressed as follows:
\begin{align}
    \y_{u} \triangleq \H_{{\rm DL},u}\P_{\rm TX}\W_{\rm TX}\V\s + \n_u,
\end{align}
where $\n_u\sim\mathcal{CN}(\mathbf{0},\sigma^2_u\mathbf{I}_L)$ denotes the Additive White Gaussian Noise (AWGN) vector. Similarly, the baseband received signal $\y\in\Compl^{N_{\rm RF}\times1}$ at the output of the RX RF chains of the considered DMA-based FD transceiver is given by:
\begin{align}\label{eq:received}
    \y \triangleq& \W^{\rm H}_{\rm RX}\P^{\rm H}_{\rm RX}\H_{\rm R}\P_{\rm TX}\W_{\rm TX}\V\s \\
    &\nonumber+ (\W^{\rm H}_{\rm RX}\P^{\rm H}_{\rm RX}\H_{\rm SI}\P_{\rm TX}\W_{\rm TX} + \D)\V\s + \W^{\rm H}_{\rm RX}\P^{\rm H}_{\rm RX}\n,
    %\\
    %\nonumber=&\W^{\rm H}_{\rm RX}\P^{\rm H}_{\rm RX}\H_{\rm R}\P_{\rm TX}\W_{\rm TX}\V\s + (\widetilde{\H}_{\rm SI}+\D)\V\s+ \W^{\rm H}_{\rm RX}\P^{\rm H}_{\rm RX}\n,
\end{align}
where $\n\sim\mathcal{CN}(\mathbf{0},\sigma^2\mathbf{I}_{N_{\rm RF}})$ represents the AWGN vector and $\H_{\rm SI}\in\Compl^{M\times N}$ is the near-field SI channel, which is defined $\forall$$i,i'=1,2,\dots,N_{\rm RF}$ and $\forall$$n,n'=1,2,\dots,N_{\rm E}$, as:
\begin{align}
    &\nonumber[\H_{\rm SI}]_{(i-1)N_E+n,(i'-1)N_E+n'} \!\triangleq\! \alpha_{i,i',n,n'}\exp\left(\frac{\jmath2\pi}{\lambda}r_{i,i',n,n'}\right)\!,
\end{align}
where $\theta_{i,i',n,n'}$, $r_{i,i',n,n'}$, and $\alpha_{i,i',n,n'}$ are defined analogous to \cite{gavras2023full} and $\D\in\mathbb{C}^{N_{\rm RF}\times N_{\rm RF}}$ is the digital SI cancellation.

\section{Proposed JCAS Framework}\label{Sec: 3}
%We herein optimize the TX/RX analog BF, the TX digital BF matrix, and the digital SI cancellation matrix to maximize DL rate while enabling simultaneous targets' parameters tracking.

\subsection{TX BF, Precoding and SI Cancellation}\label{sec: Opt}
Assuming the availability of the UE parameter estimations $(\widehat{r}_u,\widehat{\theta}_u,\widehat{\varphi}_u)$ $\forall u$ via the MUSIC-like approach in \cite{gavras2023full}, estimates of the DL channels, $\widehat{\H}_{\rm DL,u}$'s, can be composed via \eqref{eqn:DL_chan}. With this information collected at the considerd DMA-based FD transceiver, we formulate the optimization problem: 
\begin{align}
        \mathcal{OP}&:\nonumber\underset{\substack{\W_{\rm TX}, \V,\D}}{\max} \quad \sum\limits_{u=1}^{U}\Big(\|\widehat{\H}_{\rm DL,u}\P_{\rm TX}\W_{\rm TX}\V\|^2\sigma_u^{-2}\Big)\\
        &\nonumber\text{\text{s}.\text{t}.}\,
        \|[\W^{\rm H}_{\rm RX}\P^{\rm H}_{\rm RX}\H_{\rm SI}\P_{\rm TX}\W_{\rm TX}\V]_{(i,:)}\|^2\leq \gamma,\, \forall i,\\
        &\nonumber\,\quad \sum\limits_{u=1}^{U}\|\P_{\rm TX}\W_{\rm TX}\V_u\|^2 \leq P_{\rm max},\hspace{0.5em}  w^{\rm TX}_{i,n},w^{\rm RX}_{i,n} \in \mathcal{W},
\end{align}
targeting the maximization of the summation of the received Signal-to-Noise-Ratios (SNRs) at all UEs. The first constraint refers to the residual SI threshold $\gamma$ at the output of each $i$th microstrip of the RX DMA. To solve this non-convex problem with coupled variables, we apply an alternating optimization approach. For the TX DMA analog BF matrix $\W_{\rm TX}$, we first constrain its elements to the set $\mathcal{F}=\{e^{j\phi}|\phi\in[-\pi/2,\pi/2]\}$ (e.g., DFT codebook) with constant amplitude and arbitrary phase values. We then solve the following optimization problems through an one-dimensional search:

\begin{figure*}[!t]
  \begin{subfigure}[t]{0.41\textwidth}
  \centering
    \includegraphics[width=0.85\textwidth]{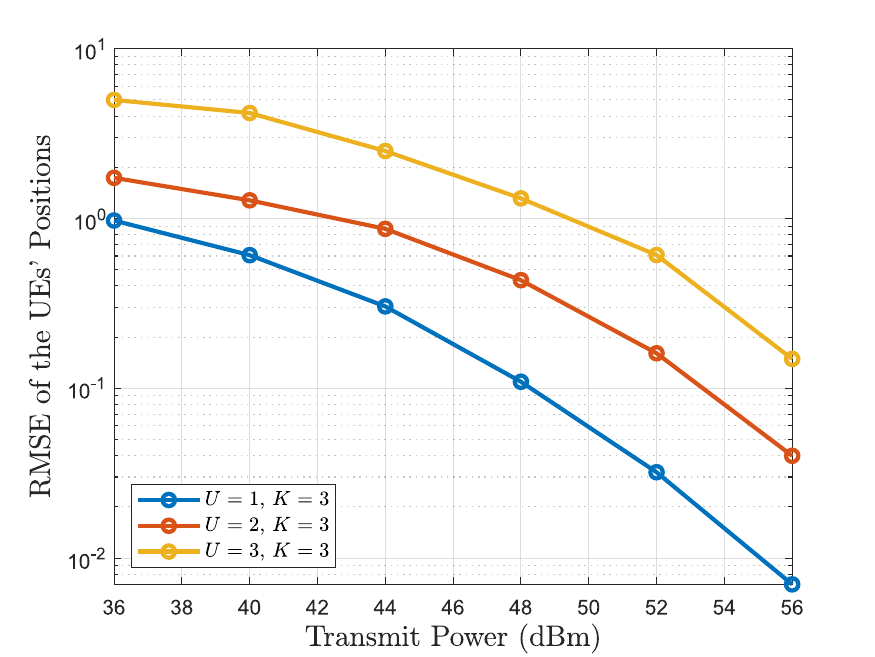}
    \caption{UEs position estimation.}
    \label{fig:RMSE}
  \end{subfigure}\hfill
  \begin{subfigure}[t]{0.45\textwidth}
  \centering
    \includegraphics[width=0.8\textwidth]{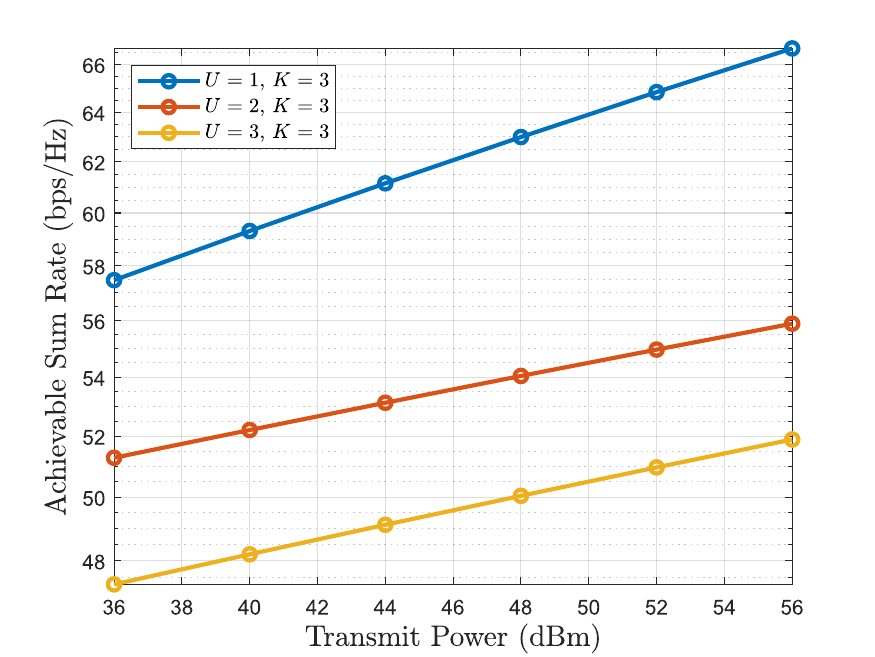}
    \caption{DL rate performance.}
    \label{fig:DL}
  \end{subfigure}
      \caption{\small{Tracking and DL sum-rate performance for $K=3$ targets, with $U=\{1,2,3\}$ out of which being the UEs each with $L=2$ antennas, versus the transmit power $P_{\rm max}$ for an FD holographic MIMO node with $N_{\rm RF} = 4$ TX/RX microstrips each with $N_{\rm E}=512$ metamaterials.}}\vspace{-0.4cm}
  \label{fig:Estimation_vs_P_T}
\end{figure*}
\begin{align*}
    \begin{array}{cc}
        \mathcal{OP}1:\underset{\widetilde{\W}_{\rm TX}}{\max} \,\, \|\widehat{\H}_{\rm R}\widetilde{\W}_{\rm TX}\|^2
        \,\,\text{\text{s}.\text{t}.}\,\, \widetilde{w}^{\rm TX}_{i,n}\in \mathcal{F},
    \end{array}
\end{align*}
where $\widehat{\H}_{\rm R}$ is constructed similar to $\widehat{\H}_{\rm DL,u}$'s based on \eqref{eq:H_R}. Given $\widetilde{\W}_{\rm TX}$ and compensating for the signal propagation inside the microstrips, the final TX DMA weights can be designed as $w^{\rm TX}_{i,n} \triangleq (\jmath+\widetilde{w}^{\rm TX}_{i,n}e^{\jmath\rho_{i,n} \beta_{i}})/2$. We finally compute the precoding matrix $\V$ and the digital SI cancellation matrix $\D$ in a manner akin to \cite{gavras2023full}.

\subsection{Target Tracking and RX Combining}
The targets existing in the system are tracked at the DMA-based RX part of the considered FD holographic MIMO transceiver, through the reception of their reflections of the DL data signals and the PASTd algorithm~\cite{774778}. This involves tracking the signal subspace $\U_s$ and its corresponding eigenvalues $\mathbf{\Lambda}_s$ in baseband. By exploiting the orthogonality between $\U_s$ and the noise subspace $\U_n$, with $\U_n\U_n^{\rm T} = \I-\U_s\U_s^{\rm T}$, \cite{gavras2023full}'s MUSIC variation is used for the PASTd initialization and the targets' parameters estimation. 

As mentioned in the previous Section~\ref{sec: Opt}, initial estimations of the UEs' parameters are needed in $\mathcal{OP}1$ to compute $\W_{\rm TX}$, while keeping a fixed wide analog combining configuration $\W_{\rm RX}$ throughout the tracking process. The latter aims to maximize the DL sum rate using a wide RX combiner to detect reflecting signals, even when UE moves. It is noted that, in the context of near-field operations, a wide RX beam is deemed mandatory for increased precision positioning. We acknowledge that optimizing the RX DMA weights based on UEs' positions may compromise performance as they move. Assuming UE mobility with position changes every $T$ Transmission Time Intervals (TTIs), one can regularly update the signal subspace and its corresponding eigenvalues. To this end, at the beginning of each communication block, we recalculate $\W_{\rm TX}$ to account for the UEs' mobility.

\section{Numerical Results}\label{sec: num}
A DMA-based FD holographic MIMO transceiver with $N_{\rm RF}=4$ microstrips and $N_{\rm E}=512$ metamaterials per microstrip, spaced at $\lambda/2$ and $\lambda/5$, respectively, is considered in this section in a simulation scenario including $K=3$ targets, with varying subsets of them being the DL UEs equipped with $L=2$ antennas. The separation between TX and RX DMAs is $d_{pl}=2d_{\rm P}=0.04$ meters, and the JCAS system operates at a central frequency of $120$ GHz with a $B=150$ KHz bandwidth. The UEs were randon placed within the Fresnel region with coordinates $\phi_u = 90^{\circ}$, $\theta_u \in [0^{\circ}, 90^{\circ}]$, and $r_u \in [1, 20]$ meters. The results that follow were obtained from independent Monte Carlo runs of $T=200$ TTIs for initial estimation and another $T=100$ for tracking. We have considered that, at every $T$ TTIs, all UEs update their position for $100$ communication blocks using the uniform distribution as $k_{new} = k + \mathcal{U}(\mu_k,d_k)$, where $k$ is a string referring to either the range or the elevation angle (i.e., $k\in\{r,\theta\}$), and $\mu$ and $d_k$ denote their mean and maximum allowable deviation, respectively. At periodic intervals of transmission blocks, we set $\mu_r$ and $\mu_{\theta}$ equal to the UEs' coordinates. This ensures smooth UE movement without causing jitter around a fixed point. The noise variances $\sigma^2$ and $\sigma_1^2$ were set to $-174 + 10\log_{10}(B)$, and a $10$-bit beam codebook $\mathcal{F}$ was utilized for the TX analog BF matrix $\W_{\rm TX}$ in $\mathcal{OP}1$.

The tracking and sum-rate performance of the proposed JCAS system is demonstrated in Figs.~\ref{fig:RMSE} and~\ref{fig:DL}, respectively, versus $P_{\rm max}$ in dBm. All UEs were assumed to move with parameters $\mu_{\theta}=\mu_r=2$, $d_{\theta}=5$, and $d_r=2$. In Fig.~\ref{fig:RMSE}, the Root Mean square error (RMSE) for all targets averaged over all TTIs is illustrated, whereas, Fig.~\ref{fig:DL} depicts the corresponding sum-rate performance. As expected, both metrics improve with increasing SNR values. It is also shown that the proposed estimation scheme is robust in tracking multiple UEs, even with the considered small number of RX RF chains. However, as the number of UEs increases, localization gets degraded, which also deteriorates the achievable sum rate. The latter happens because the TX design relies on the composition of the DL channel via the estimated UEs' coordinates.

\section{Conclusion}
We proposed a JCAS system constituting of a DMA-based FD holographic MIMO transceiver and presented a scheme for simultaneous multi-user communications and target tracking in the near-field regime. Our numerical results showcased that this dual functionality performance depends on the number of nodes in the system as well as the number of RX RF chains.  

\newpage
\bibliographystyle{IEEEtran}
\bibliography{IEEEabrv,ms}
\end{document}